\documentstyle[amssym,emulateapj]{article} 

\newcommand{\figa}{1}
\newcommand{\figb}{2}
\newcommand{\figc}{3}
\newcommand{\figd}{4}
\newcommand{\fige}{5}

\def\kms{km~s$^{-1}$}
\def\cm2{cm$^{-2}$}

\def\lya{{\rm Ly}$\alpha$}
\def\lyb{Ly$\beta$}

\newcommand{\lsim}{\ \raise -2.truept\hbox{\rlap{\hbox{$\sim$}}\raise5.truept
        \hbox{$<$}\ }}
\newcommand{\gsim}{\ \raise -2.truept\hbox{\rlap{\hbox{$\sim$}}\raise5.truept
        \hbox{$>$}\ }}

\begin{document}

\input{epsf}
\input{psfig} 

\title{THE LYMAN--$\alpha$ FOREST OF THE QSO IN THE HUBBLE DEEP FIELD
SOUTH\altaffilmark{1}}

\altaffiltext{1}{Based on observations made with the NASA/ESA {\sl
Hubble Space Telescope} by the Space Telescope Science Institute,
which is operated by AURA, Inc., under NASA contract NAS 5--26555.}

\author{S. Savaglio \altaffilmark{1,2},
H. C. Ferguson \altaffilmark{1},
T. M. Brown \altaffilmark{3}, 
B. R. Espey \altaffilmark{1,2},
K. C. Sahu \altaffilmark{1},
S. A. Baum \altaffilmark{1},
C. M. Carollo \altaffilmark{4},
M. E. Kaiser \altaffilmark{4},
M. Stiavelli \altaffilmark{1,2,5},
R. E. Williams \altaffilmark{1},
J. Wilson \altaffilmark{1}
}

\affil{\altaffilmark{1}
Space Telescope Science Institute, 3700 San Martin Drive, 
	Baltimore, MD21218, USA}

\affil{\altaffilmark{2} 
On assignment from the Space Science Department of the European
Space Agency}

\affil{\altaffilmark{3} 
Laboratory for Astronomy \& Solar Physics, 
Code 681, NASA/GSFC, Greenbelt, MD 20771}

\affil{\altaffilmark{4} 
Johns Hopkins University, 3701 San Martin Dr., 21218 MD, USA}

\affil{\altaffilmark{5}
Scuola Normale Superiore, Piazza dei Cavalieri 7, I56126, Pisa, Italy}

\begin{abstract}
The quasar in the Hubble Deep Field South (HDFS), J2233--606
($z_{em}=2.23$) has been exhaustively observed by ground based
telescopes and by the STIS spectrograph on board the Hubble Space
Telescope (HST) at low, medium and high resolution in the spectral
interval from 1120 \AA~to 10000 \AA. The combined data give continuous
coverage of the Lyman--$\alpha$ forest from redshift 0.9 to 2.24. This
very large base--line represents a unique opportunity to study in
detail the distribution of clouds associated with emitting structures
in the field of the quasar and in nearby fields already observed as
part of the HDFS campaign.  Here we report the main properties
obtained from the large spectroscopic dataset available for the
\lya~clouds in the intermediate redshift range $1.20-2.20$, where our
present knowledge has been complicated by the difficulty in producing
good data.  The number density is shown to be higher than what is
expected by extrapolating the results from both lower and higher
redshifts: $63\pm8$ lines with $\log N_{HI}\geq14.0$ are found
(including metal systems) at $<z>=1.7$, to be compared with $\sim 40$
lines predicted by extrapolating from previous studies.  The redshift
distribution of the Lyman--$\alpha$ clouds shows a region spanning
$z\simeq 1.383-1.460$ (comoving size of 94 $h^{-1}_{65}$ Mpc,
$\Omega_o=1$) with a low density of absorption lines; we detect 5
lines in this region, compared with the 16 expected from an average
density along the line of sight.  The two point correlation function
shows a positive signal up to scales of about 3 $h^{-1}_{65}$ Mpc and
an amplitude that is larger for larger HI column densities. The
average Doppler parameter is about 27 \kms, comparable to the mean
value found at $z > 3$, thus casting doubts on the temperature
evolution of the \lya~clouds.
\end{abstract}


\section{INTRODUCTION}

A satisfactory connection between quasar absorption lines and galaxies
close to the QSO line of sight has been provided so far only by
relatively low redshift studies ($z<1.5$), as galaxies at high
redshifts are much harder to observe.  For $z>2$, large samples of
quasar absorption lines have been provided because at these redshifts
the many strong UV lines are redshifted to the optical range, where
very sensitive ground based observations can be obtained. However a
complete picture in the whole observable redshift range is missing
because of the lack of information at intermediate redshifts ($1.5 < z
< 2$).

With the advent of the high resolution UV spectrograph Space Telescope
Imaging Spectrograph (STIS), a detailed study of the intermediate
redshift \lya~forest and the connection to galaxy properties can now
be obtained. Indeed, this was one of the motivations for the Hubble Deep
Field South (HDFS). Much of the ground-based
and HST HDFS campaign (Williams et al.\ 1999) has been
devoted to the observations of the quasar J2233--606 ($z_{em}=2.23$)
that lies in the STIS field (about 5 and 8 arcmin from the WFPC2 and
NICMOS fields).  The study of the quasar incorporates both
spectroscopy (Sealy et al.\ 1998; Savaglio 1998; Outram et
al.\ 1998; Ferguson et al.\ 1999) and imaging (Gardner et al.\ 1999).

Here we present the interesting features shown by the distribution of
the \lya~clouds obtained by the fitting of the absorption lines in the
redshift range $1.20 < z < 2.20$ in the high and medium resolution
spectra taken with STIS/HST (Ferguson et al.\ 1999\footnote{See
also the HDFS web site at
http://www.stsci.edu/ftp/observing/hdf/hdfsouth/hdfs.html}) and
UCLES/AAT (Outram et al.\ 1998). The spectral resolution ranges from
FWHM = 10 \kms~in the interval $\lambda\lambda=2670-3040$~\AA, to 50
\kms~in $\lambda\lambda=3040-3530$~\AA, and to 8.5 \kms~in
$\lambda\lambda=3530-3900$~\AA.

\section{The Doppler parameter and the column density redshift
distribution}

The line parameters have been obtained using the MIDAS package
FITLYMAN (Fontana \& Ballester 1995) in the spectral range 2670--3900
\AA~through $\chi^2$ minimization of Voigt profiles.  At
shorter wavelengths, the Lyman limit of the system at $z\simeq1.943$
absorbs a large fraction of the QSO flux, preventing the fit of
absorption lines.  The optical UCLES/AAT and STIS/HST spectra have
been combined to simultaneously fit the Lyman series of the Ly$\alpha$
clouds and that gives more robust results than using the Ly$\alpha$
line alone.  Only the parameters of \lya~clouds at $z<1.60$ have
been obtained using the \lya~absorption line alone.  The redshift
distribution of the Doppler parameters and of the column densities are
shown in Fig.~\figa, together with the instrumental Doppler width
and column density 4$\sigma$ detection limit for the \lya~lines.  The
sample does not include HI column densities associated with known
metal systems, and lines close ($z>2.2$) to the quasar redshift that
are considered to be affected by the ionization of the quasar.
The number of identified intervening metal systems is 5. Three of them
have a complex structure (for a total of 10 \lya~components) with HI
column density in each system $\log N_{HI}>16$. These clouds are
presumably associated with galaxies and we consider them a different
population of clouds with respect to lower HI column density clouds of
the intergalactic medium, although many of them are presumably
polluted of metals by nearby galaxies (Cowie \& Songaila, 1998). Metal
systems have been included in the sample only to compare the number
density redshift evolution with previous studies at lower and higher
redshifts (see section 3).  The total number of \lya~clouds excluding
metal systems is then 210. The reported 4$\sigma$ detection limit
is determined by considering the error array of the spectra and does
not take into account the line blending effect; at $z>1.7$ it is not
very important because the line density at intermediate redshifts is
relatively low, but it becomes particularly strong at lower redshifts,
due to the presence of Lyman series lines of the higher redshift
\lya~clouds that prevents detection of weak lines ($N_{HI} <
10^{13.5}$ \cm2). We therefore consider the completeness limit in the
whole redshift range of $N_{HI} \geq 10^{14}$ \cm2.

\vskip 1em \hskip -1.1em
\parbox{3.25in}{\epsfxsize=8.8cm \epsfysize=8.8cm \epsfbox{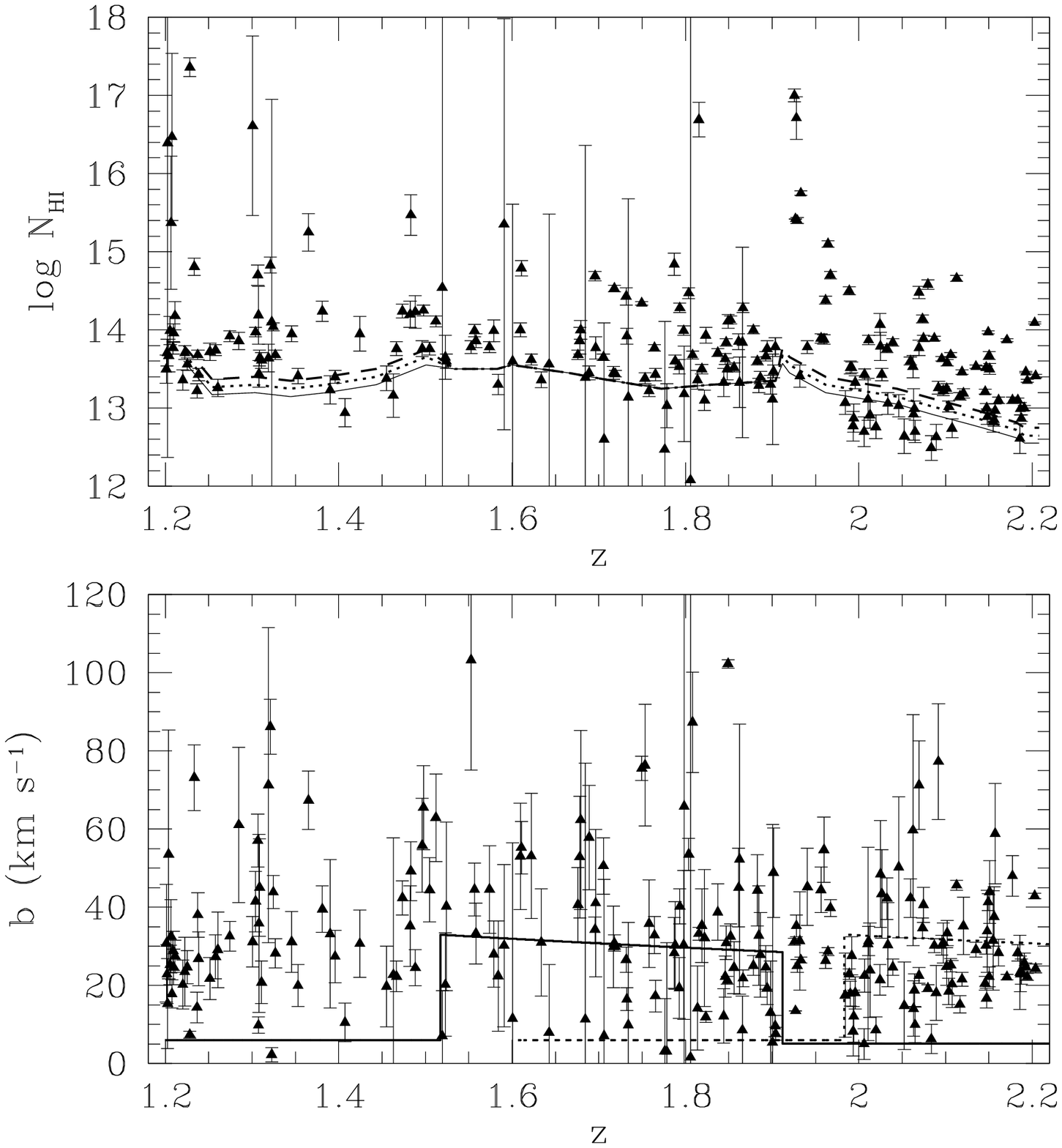}}
\centerline{\parbox{3.35in}{\small {\sc Fig.~\figa--}
\hskip -.1em
The Doppler parameter (lower panel) and column
density (upper panel) vs. redshift of the \lya~forest along
the J2233--606 line of sight. The solid and dotted lines in the lower panel
are the instrumental Doppler widths along the spectra for the \lya~
and \lyb~forests respectively.  The dashed, the
dotted and the solid lines in the upper panel represent the 4$\sigma$ HI
column density detection limit in the case of Doppler parameter of 40,
30 and 20 \kms~respectively.}
}
\vskip 1em

Some interesting features can be noticed in the column density
redshift distribution. There is a peak of high HI \lya~clouds at
$z=1.92-1.99$ ($\sim60$ $h^{-1}_{65}$ Mpc comoving). A metal system
with a $z\simeq 1.943$ Lyman limit was first detected from the test
STIS observations of the quasar. Therefore this peak might indicate
the presence of a high density of galaxies at those redshifts. At
lower redshifts (in the interval $1.383 < z < 1.460$), we see a region
with a low density of lines.  The corresponding line density is
$dn/dz=65$, compared with an observed mean over the whole range of
$dn/dz=210$ lines. There are no lines with HI column density larger
than the completeness limit of 10$^{14}$ atoms/\cm2, while from the
mean observed in the whole redshift range we would expect 4 lines.
Although this is not statistically very significant, it is suggestive
of the presence of a ``void'' of comoving size of 94 $h_{65}^{-1}$
Mpc.  The redshift measurements of galaxies from multi--color ground
observations and from the very deep STIS images of the field will
probably confirm if this is real. The lack of \lya~lines can also be
caused by a strong ionization of the clouds by the local UV radiation
field, for instance in a quasar environment.

It is worth noting the decrement of lines going from the QSO redshift
to $\sim 1.38$, shown by the histogram of the entire sample
(Fig.~\figb). This effect disappears when selecting strong lines,
and so it may be due to the detection limit variation along the
spectrum.  For smaller redshifts, the number of lines increases
again. We notice that at $z\simeq1.335$, a CIV system has been
tentatively identified (Ferguson et al.\ 1999) and a quasar at a
distance of about 44.5" from the line of sight ($\sim 300$
$h_{65}^{-1}$ kpc) has been found from the ground by EMMI/NTT
observations (Tresse et al. 1998).  This might also be an indication of 
an overdensity of objects at that redshift.

\vskip 1em
\hskip -1em
\parbox{3.25in}{\epsfxsize=9cm \epsfysize=9cm \epsfbox{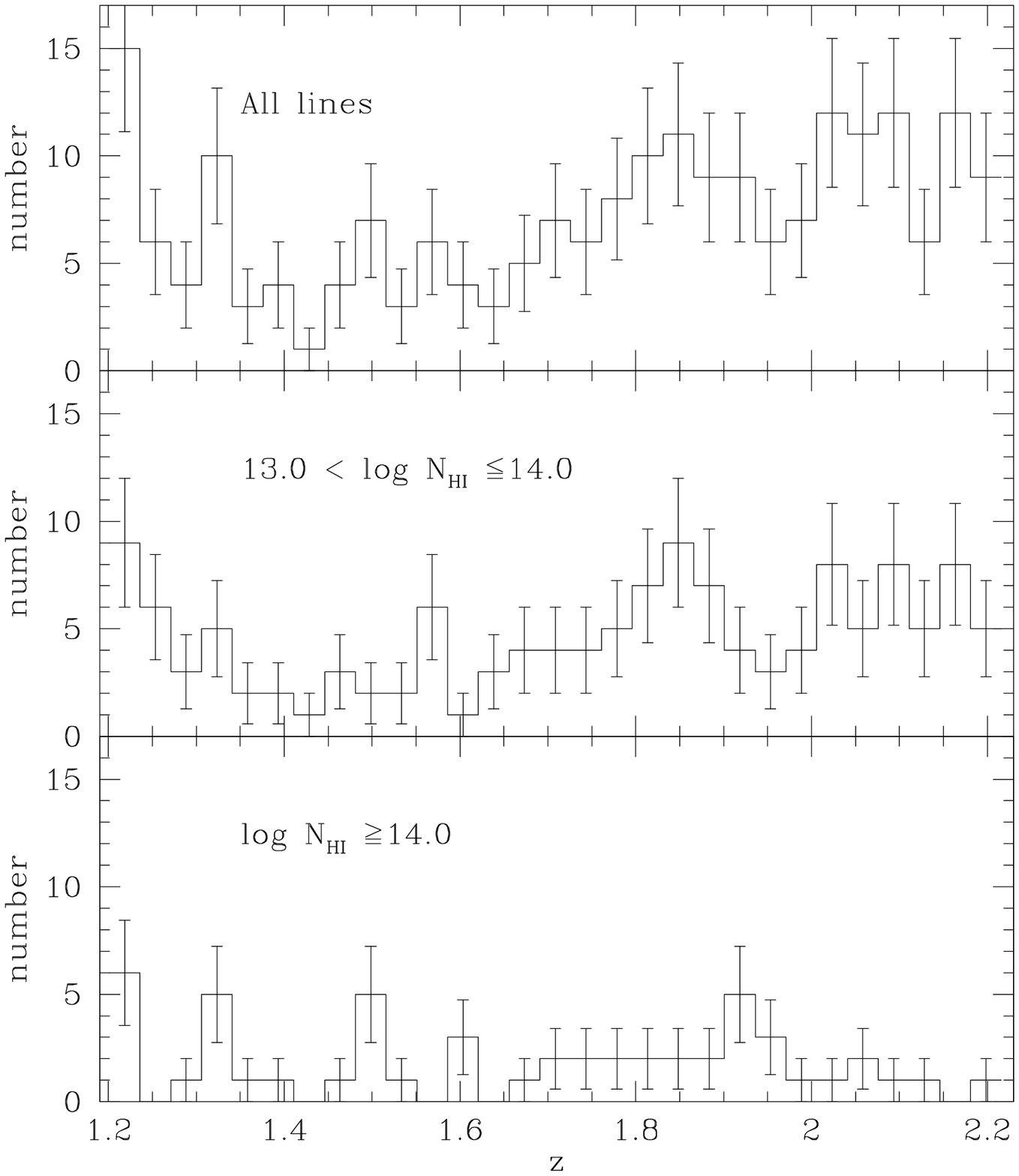}}
\vskip 0.1in 
\centerline{\parbox{3.35in}{\small {\sc Fig.~\figb--}
Histogram of the HI column density of \lya~clouds 
as a function of redshift for different thresholds. Error bars are the
square-root of the number in each bin.}
}

\section{The number density evolution at $z=1.20-2.20$}

The number density evolution of the \lya~forest, described by a power
law of the type $dn/dz \propto (1+z)^\gamma$, has been studied at high
redshift ($1.7 < z < 4.1$) using samples of high resolution data by
Giallongo et al.\ (1996) and Kim et al.\ (1997).  For lines with $\log
N_{HI} \geq 14.0$, $dn/dz$ shows a fast evolution, with $\gamma\simeq
3.6$. The analysis at low redshifts (Weymann et al.\ 1998), based on low
resolution spectroscopy of the Quasar Absorption Line Survey (QALS), gives in
the range $0.0 < z < 1.5$ a much flatter $dn/dz$, consistent with weak
negative evolution (for $\Omega_o=1$) $\gamma=0.16\pm0.16$.  We note
that in the two redshift regimes a variation of $\gamma$ with column
density threshold has been found (larger for increasing threshold).
This would go the opposite direction than what is shown in
Fig.~\figb\ where $\gamma$ is smaller for higher threshold. The
discrepancy can be explained if a large number of lines with
$\log N_{HI} <14$ is present, but not detected at $z=1.4-1.8$.

\vskip -1em \hskip -1em
\parbox{3.25in}{\epsfxsize=8.7cm \epsfysize=8.7cm \epsfbox{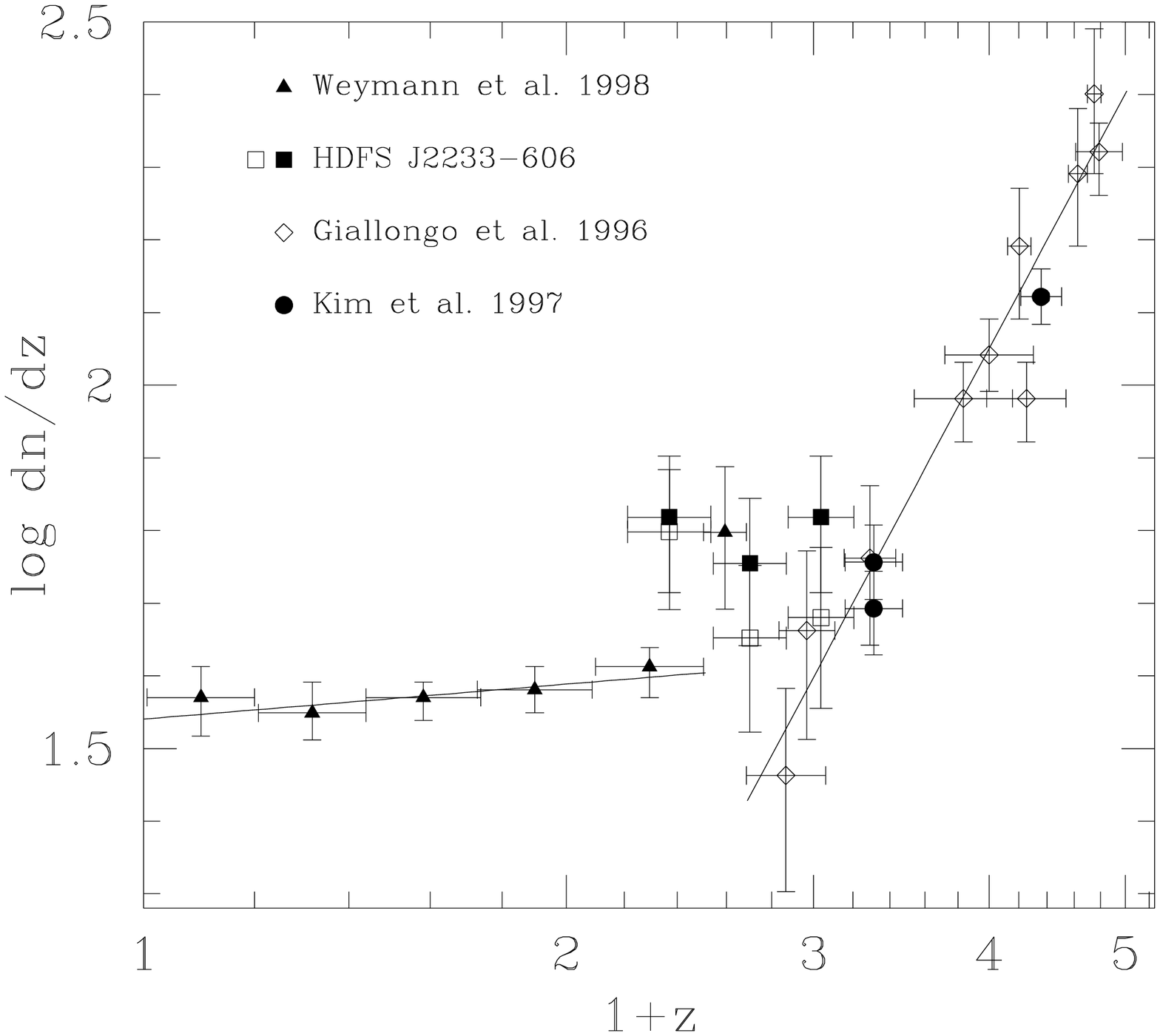}}
\vskip -0em
\centerline{\parbox{3.35in}{\small {\sc Fig.~\figc--} Number density
evolution of the \lya~clouds with $\log N_{HI}\geq14.0$ from $z=0$ to
$z=4$. Filled symbols are for samples that include metal systems,
open symbols do not. In the sample of Kim et al., only lines with 
$\log N_{HI}\leq16.0$ are included. The low redshift line is the fit with
$\gamma=0.16$ obtained by a sample of \lya~clouds with equivalent
width $EW\geq0.24$ \AA~(Weymann et al.\ 1998) which corresponds to
$\log N_{HI}\geq 14.0$ for $b\sim 26$ \kms. The high redshift line is
the fit with $\gamma=3.62$.}
}

\vskip 1em

The problem in comparing high and low redshift data stems from the
different techniques used to measure column densities. At high
redshifts, high resolution observations allow line fitting of Voigt
profiles, and those give more robust results than the
curve--of--growth technique used to estimate HI from equivalent width
measurements of low resolution data at low redshifts. On the other
hand, the strong evolution of line density makes the line deblending
of complex structures rather difficult at high redshifts. The
\lya~forest of J2233--606 lies between the two redshift regimes and so
it is extremely useful to evaluate the correct connection between the
two.  The results for the two samples of \lya~clouds with and
without metal lines are shown in Fig.~\figc\ for lines with $\log
N_{HI} \geq 14.0$.  The data from J2233--606 have been binned into
three redshift ranges $\Delta z = 1.201-1.535,1.535-1.870$ and
$1.870-2.204$ corresponding roughly to the STIS high resolution and
medium resolution intervals and to the UCLES high resolution
interval. The last two bins give a number density of lines that would
better match the high and low redshift observations if metal systems are
excluded.  The first bin gives a larger number of lines,
$66\pm14$ including metal systems, compared with $40\pm4$ lines given
the QALS best fit (Weymann et al.\ 1998).  That deviation is not
particularly significant, and could indicate a peculiarity along the
line of sight to J2233--606. However, a similar excess at $1.505 < z <
1.688$ ($63\pm14$) is also shown by the quasar UM18, which is part of
the FOS sample of the QALS, but not included in the fitting of the
number density evolution.  The excess found in J2233--606 can be due
to the presence of a cluster of lines at $z\sim1.335$, as suggested in
the previous section.  Other possible explanations would be the
blending effect by which nearby lines with column density below the
threshold would artificially appear in the limited signal--to--noise
spectrum as stronger single lines with HI above the threshold, or the
presence of a small number ($\sim5$ would be enough) of unidentified
metal lines interpreted as \lya~lines.  This latter possibility looks
very unlikely because even in clusters, metal lines typically appear
very narrow at high resolution.

\vskip -1em
\hskip -1em
\parbox{3.25in}{\epsfxsize=9cm \epsfysize=9cm \epsfbox{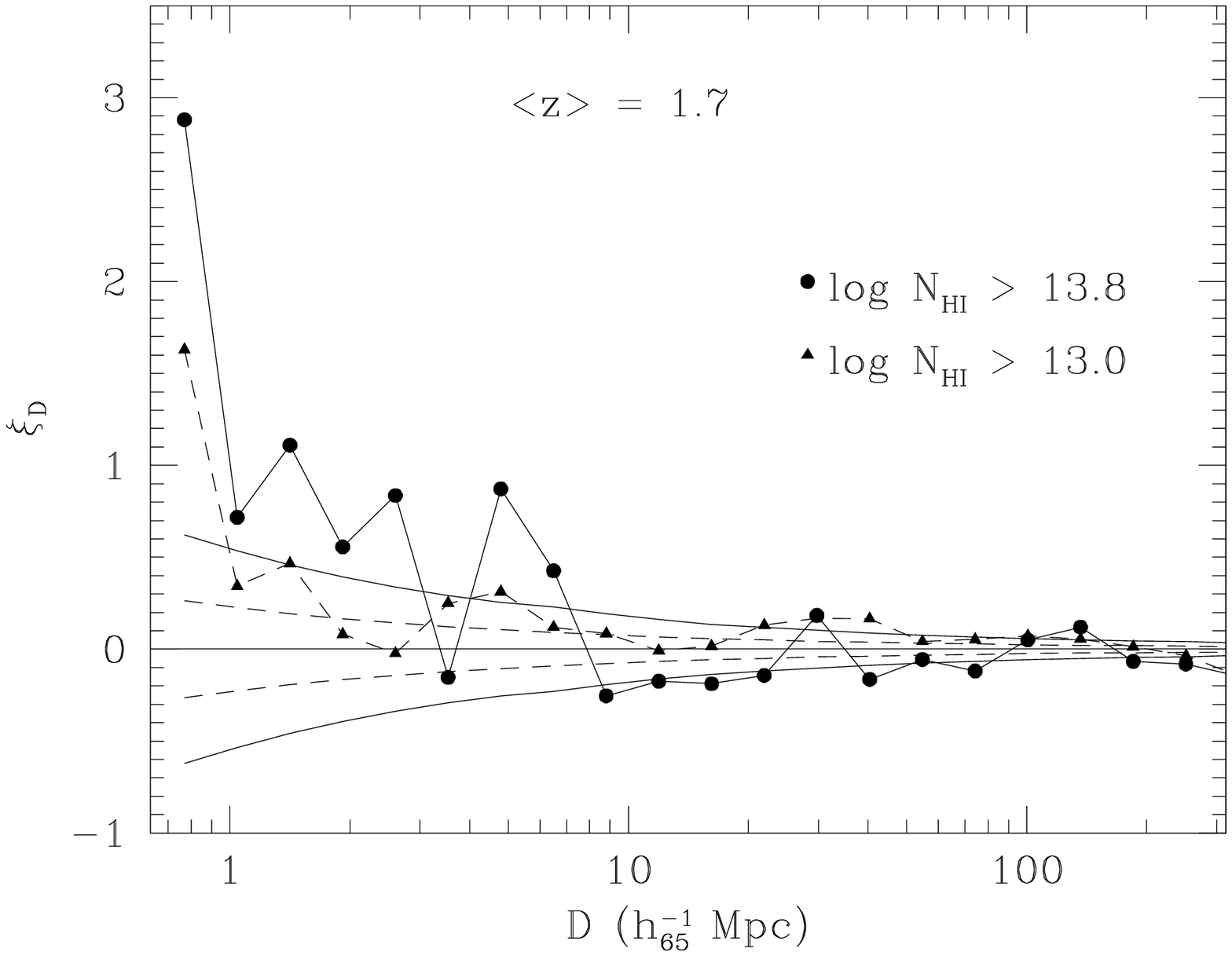}}
\vskip -0.65in
\centerline{\parbox{3.35in}{\small {\sc Fig.~\figd--}
Two point correlation function of the \lya~forest  in 
J2233--606 for lines with $\log N_{HI}>13.0$ and 13.8. The Poisson error
is also given as dashed and solid lines respectively.}
}

\section{The clustering properties}

A positive signal in the clustering properties of the \lya~forest has
been found thanks to the advent of high resolution spectroscopy. The
investigation has been mainly based on the analysis of the two point
correlation function (Cristiani et al.\ 1997; Kim et al.\ 1997) but
also on that of the power spectrum of mass density fluctuations (Hui
et al.\ 1997; Croft 1998; Amendola \& Savaglio 1998).  Results have
shown that the clustering is present up to $\Delta v \sim 300$
\kms~and that the amplitude increases with increasing column density
threshold of the lines. Most interesting for models of large scale
structure formation is the redshift evolution found, showing larger
values at smaller redshifts. Although the analysis of the two point
correlation function is uncertain when the sample is too small, we
report the result on the forest of J2233--606 in Fig.~\figd\ for two
column density thresholds. The signal is significant in the first bin
($D\sim0.8$ $h^{-1}_{65}$ Mpc) and it is higher at the higher HI
threshold at about the 6 and 4 $\sigma$ level for $\log N_{HI}>13.0$
and 13.8 respectively. In a previous study by Cristiani et al.\
(1997), the two point correlation function for $\log N_{HI}>13.8$ at
$\Delta v = 100$ \kms~has shown a redshift evolution, being
$\xi_{100}\simeq0.20$ at $<z>=3.85$ (consistent with no clustering),
$\xi_{100}\simeq0.75$ at $<z>=3.40$, and $\xi_{100}\simeq0.85$ at
$<z>=2.40$. The line of sight of J2233--606 shows 
$\xi_{100}\simeq1.3$ at $<z>=1.7$ and confirms this evolution.

\section{The Doppler parameter distribution}

Thanks to the high resolution observations of J2233--606, the analysis
of the distribution of the Doppler parameter of the \lya~forest can be
performed for the first time at these low redshifts.  At higher
redshifts Kim et al.\ (1997) have found that the median value of the
Doppler parameter is $b\sim30$ \kms~at a redshift of $z\sim3.7$, and
appeared to increase to $b\sim35-40$ \kms~at a redshift of $z\sim2.3$ for
lines with $13.8 < \log N_{HI} < 16$, which has been interpreted as an
increase of the temperature or kinematic broadening of the clouds. The
Doppler distribution in J2233--606 is shown in Fig.~\fige\ for the
whole sample and for the subsample of lines (60\% of the total) with
smaller uncertainties ($\sigma(b)<8$ \kms~and $\sigma(\log N_{HI})<
0.5$).  The fit of the two Gaussians (in the interval $0<b<60$ \kms)
gives in the two cases a mean value of 27 and 26 \kms~respectively
($\chi^2$ is 2.4 and 1.5), and these are slightly different than the
median value (29 and 25 \kms) in the whole $b$ range. The median
values are larger if lines with $13.8 < \log N_{HI} < 16$ are
selected, being 31 \kms~ in the two cases. Thus, the median value we
find over the redshift range $1.20 < z < 2.20$ is considerably below
the value of $b\sim40$ \kms~expected on the basis of the redshift
evolution suggested by Kim et al.\ (1997).

\section{Conclusions}

In contrast to the empty field used for the HDF North program, the
HDFS has been selected close to a high redshift quasar, J2233--606
($z_{em}=2.23$).  The QSO is centered in the STIS field, located
approximately 5 and 8 arcmin from the WFPC2 and NICMOS fields
respectively.  The \lya~forest of J2233--606 has been observed at high
and medium resolution (FWHM = 10, 50 and 8.5 \kms~ in the wavelength
intervals $\lambda\lambda=2670-3040$ \AA, $\lambda\lambda=3040-3530$
\AA~ and $\lambda\lambda=3530-3900$ \AA) using STIS/HST (Ferguson et
al.\ 1999) and UCLES/AAT (Outram et al.\ 1998), giving the opportunity
to analyze the properties of the intergalactic medium in the large
redshift interval $z=1.20-2.20$. This is the first time that the
\lya~forest is studied in detail for $z\lsim2$. More than 200
\lya~cloud redshifts have been found and 63 have HI column density in
excess of $10^{14}$ atoms/\cm2, 11 of which have associated metal
absorption.  The redshift distribution of the lines has shown a
deficiency of lines (5 are found over a mean expected of 16) in the
interval $1.383 < z <1.460$, corresponding to a comoving size of 94
$h^{-1}_{65}$ Mpc ($\Omega_o=1$). The region around the metal system
at $z=1.942$, also responsible of the Lyman limit observed for
$\lambda <2700$ \AA, shows an excess of strong lines in the interval
$1.92 < z < 1.99$, indicating a clustering of \lya~clouds. An excess
of lines is also found for $1.20 < z < 1.38$. The evolution of the
number density in the total sample shows a rapid decrease of lines
with redshift.  This is partially due to the blending effect of low HI
column density lines, which is stronger at lower redshifts. Indeed it
tends to disappear when selecting only strong lines ($N_{HI} \geq
10^{14}$ \cm2).  On average the number of strong \lya~clouds per unit
redshift is higher than the extrapolation from studies at lower and
higher redshifts: $63\pm8$ at $<z> = 1.702$ including metal systems
are found to be compared with 41 and 27 expected from the low and the
high redshift predictions. A larger sample of lines will confirm if
this line density is typical at that redshift or if it is peculiar to
the J2233--606 line of sight.  The two point correlation function has
shown a positive signal up to a scale of the order of 3 $h^{-1}_{65}$
Mpc. The amplitude found seems to confirm the evolution with redshift,
being larger at lower $z$.  The mean Doppler parameter of the
\lya~forest is around $26-27$ \kms, corresponding to a temperature, in
the case of thermal broadening, of $\sim4.3\times10^4$ K. The median
Doppler parameter evolution reported by Kim et al.\ (1997) is not
confirmed by this \lya~forest.

\acknowledgments 
It is a pleasure to thank L. Amendola, B. Carswell, S. Casertano, G. Ganis,
B. Jannuzi, M. Livio, P. Outram and R. Weymann for useful suggestions.

\vskip 2em
\hskip -1em
\parbox{3.25in}{\epsfxsize=9cm \epsfysize=9cm \epsfbox{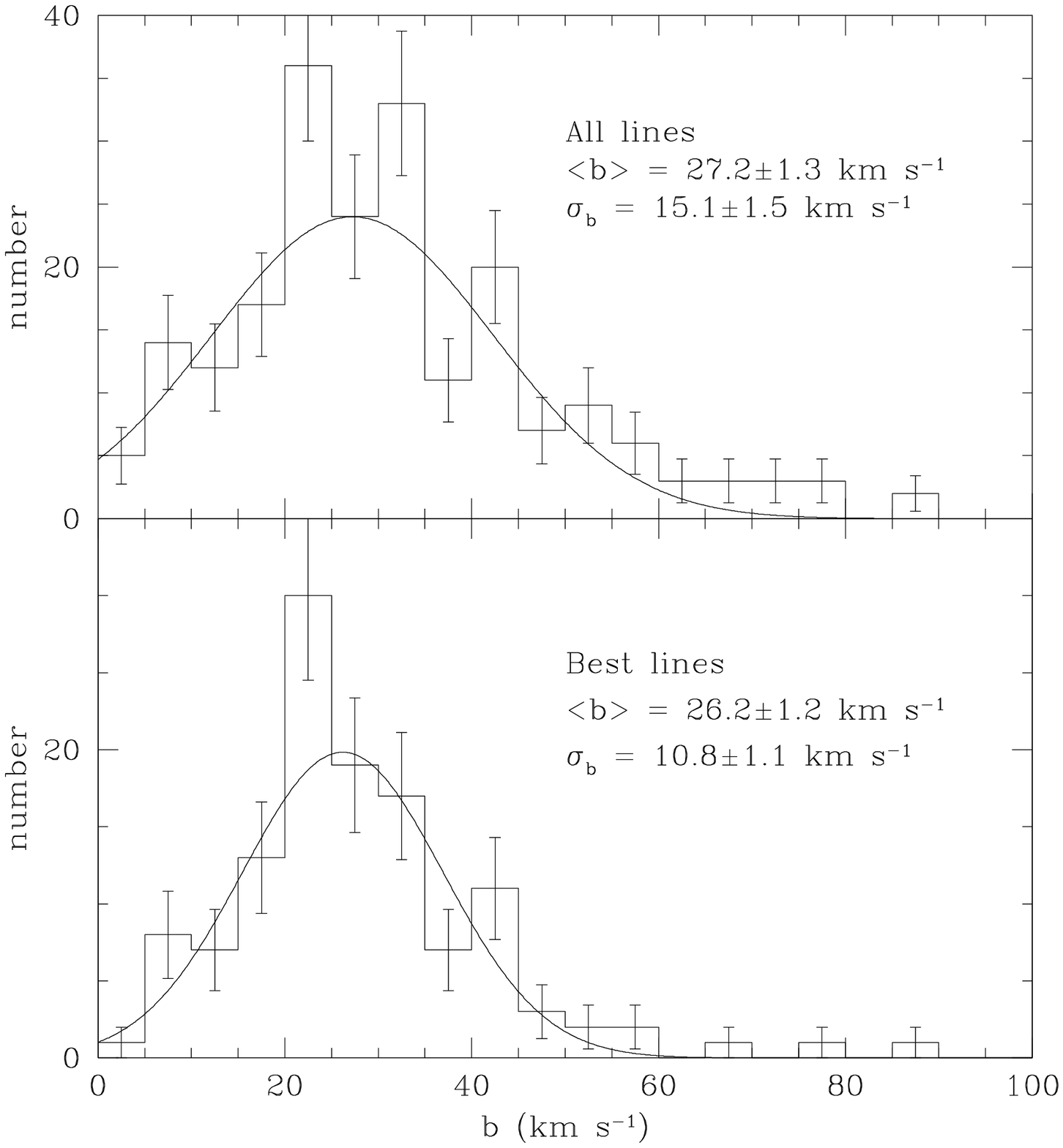}}
\centerline{\parbox{3.35in}{\small {\sc Fig.~\fige--}
Doppler parameter histograms of \lya~clouds in 
J2233--606 in the whole sample (upper panel) or in the subsample 
of lines with 1$\sigma$ error in the column density and in $b$ smaller
than 0.5 $dex$ and 8 \kms~respectively (124 lines over a total of 210). 
The fit of a 
Gaussian for $b<60$ \kms~is shown as well. Error bars are the square-root
of the number of lines in each bin.}
}

\end{document}